\documentclass[conference,letterpaper]{IEEEtran}

\usepackage[utf8]{inputenc}
\usepackage[T1]{fontenc}
\usepackage{ifthen}
\usepackage{url} % Load early as other packages might use it
\usepackage{xcolor} % For colors, often needed for TODOs or highlighting
\usepackage{calc}   % Needed for \settowidth (and general calc functions)
\usepackage[cmex10]{amsmath} % Ensure compliance with IEEE Xplore
\usepackage[normalem]{ulem} % For underlining (used by \sout)
\usepackage{enumitem} % For changing enumeration types

\usepackage{graphicx} % For including images
\usepackage{wrapfig}  % For wrapping text around figures
\usepackage[caption=false,font=normalsize,labelfont=sf,textfont=sf]{subfig} % For subfigures, caption=false is good.

\usepackage{tabularx} % For tabularx environment
\usepackage{booktabs} % For better looking table rules (optional)

\usepackage{amsfonts} % For math symbols
\usepackage{array}    % For enhanced array/tabular features
\usepackage{textcomp} % For some special characters
\usepackage{stfloats} % For better float placement (e.g. bottom-of-page floats)
\usepackage{verbatim} % For verbatim environment
\newcommand{\para}[1]{\smallskip\noindent\textbf{{#1\,}}}

\usepackage{caption} % For general caption customization
\usepackage[backend=biber,style=ieee,citestyle=numeric-comp,sortcites=true,maxnames=10]{biblatex}
\DeclareFieldFormat[online]{title}{\mkbibquote{#1\addperiod}\addspace}

\AtBeginBibliography{%
  \footnotesize % Set the desired font size for the bibliography
  \settowidth{\bibhang}{\printfield{labelprefix}99\printfield{labelsuffix}\hspace*{0.5em}}%
}

\usepackage{hyperref}
\usepackage{cleveref}
\usepackage[dvipsnames,svgnames]{xcolor} % Load xcolor with more color names
\hypersetup{
    colorlinks=true,
    linkcolor=ForestGreen,
    citecolor=ForestGreen,
    urlcolor=RoyalBlue,
    filecolor=IndianRed,
    pdftitle={Optimized Memory Tagging on AmpereOne Processors},
    pdfauthor={Shiv Kaushik, Mahesh Madhav},
    pdfsubject={MTE Architecture and Performance Analysis},
    pdfkeywords={Memory Tagging, ARM MTE, AmpereOne, Cloud Native},
    pdfnewwindow=true,
    pdfdisplaydoctitle=true,
    bookmarksopen=true
}

\begin{document}

\title{\LARGE{\textbf{Optimized Memory Tagging on AmpereOne$^\circledR$ Processors
%Cloud-Native Processor
}}}

\author{%
%  \IEEEauthorblockN{ISCA 2026 Industry Submission \#66 – Confidential Draft – Do NOT Distribute!! \\ ~}
  \IEEEauthorblockN{Shivnandan Kaushik, Mahesh Madhav, Nagi Aboulenein, Jason Bessette, Sandeep Brahmadathan,}
  \IEEEauthorblockN{Benjamin Chaffin, Matthew Erler, Stephan Jourdan, Thomas Maciukenas, Ramya Jayaram Masti,}
  \IEEEauthorblockN{Jon Perry, Massimo Sutera, Scott Tetrick, Bret Toll, David Turley, Carl Worth, Atiq Bajwa}

  \IEEEauthorblockA{ Ampere Computing, Portland, OR and Santa Clara, CA}
}

\maketitle

%\IEEEpubid{TBD 0000--0000/00\$00.00~\copyright~2021 IEEE}

\maketitle

\begin{abstract}
Memory-safety escapes continue to form the launching pad for a wide range of security attacks, especially for the substantial base of deployed software that is coded in pointer-based languages such as C/C++. Although compiler and Instruction Set Architecture (ISA) extensions have been introduced to address elements of this issue, the overhead and/or comprehensive applicability have limited broad production deployment. The Memory Tagging Extension (MTE) to the ARM AArch64 Instruction Set Architecture is a valuable tool to address memory-safety escapes; when used in synchronous tag-checking mode, MTE provides deterministic detection and prevention of sequential buffer overflow attacks, and probabilistic detection and prevention of exploits resulting from temporal use-after-free pointer programming bugs.

The AmpereOne$^\circledR$ processor, launched in 2024, is the first datacenter processor to support MTE. Its optimized MTE implementation uniquely incurs no memory capacity overhead for tag storage and provides synchronous tag-checking with single-digit performance impact across a broad range of datacenter class workloads. This paper analyzes the complete hardware-software stack of those workloads, identifying application memory management as the primary remaining source of overhead and highlighting clear opportunities for software optimization. The combination of an efficient hardware foundation and a clear path for software improvement makes the MTE implementation of the AmpereOne$^\circledR$ processor highly attractive for deployment in production cloud environments.
\end{abstract}

%\begin{IEEEkeywords}
%    Memory Tagging, ARM MTE, Memory Safety, Cloud Native
%\end{IEEEkeywords}

\IEEEpubidadjcol % This directive must be in the second column of text on the first page, so that it doesn't overlap the footer

\section{Introduction} % 1 pages

Memory safety continues to be a challenge for pointer based languages such as C/C++, where exploits of memory safety escapes provide a launching pad for a broad range of attacks~\cite{eternal_war}. The Google Chromium Project's analysis of critical and high severity security issues reported since 2015 attributed over 70\% of these to memory safety, roughly evenly split between \emph{spatial} exploits such as buffer overflow and \emph{temporal} bugs such as pointer use-after-free~\cite{chromium_vulnerability}. Similar supporting data published for Microsoft products in a 2019 BlueHat presentation and for Google Android - attributed over 70\%~\cite{microsoft_vulnerability} and 60\%~\cite{android_vulnerability} of vulnerabilities, respectively, to memory-safety escapes. While memory-safe languages such as Rust \cite{RustProgrammingLanguage, RustGitHub} are starting to gain adoption, the transition to these is slower than desired and memory safety remains a challenge for the broad base of C/C++ based software deployed in production environments \cite{MarkRussinovichRust, LinusZDNetRust, google_memory_safety_blog}. 

Software techniques - such as address space layout randomization~\cite{aslr} and stack canaries~\cite{canary} - for mitigating the impact of attacks exploiting memory-safety escapes  have long been known and widely used. Similarly, hardware features such as Intel Control Flow Enforcement Technology (CET) \cite{Intel_ROPCET} for the x86 architecture and Pointer Authentication (PAC), Branch Target Identification (BTI) \cite{ARM_PACBTI} for the ARM AArch64 architecture  are gaining traction. However, memory-safety violation detection and prevention solutions suitable for large-scale deployment in production environments have remained elusive. 

Specifically, software techniques such as Address Sanitizers in compilers (LLVM, GCC) and in the Linux Kernel have been developed for identifying memory-safety escapes during code development and testing~\cite{ASAN_llvm, ASAN_gcc, Linux_KASAN}. Historically, the x86 (Intel Memory Protection Extensions (MPX)~\cite{Intel_MPX}) and SPARC  (Application Data Integrity (ADI)~\cite{LinuxKernel_SPARC_ADI, Google_MemorySafety}) architectures have also introduced ISA extensions to detect/prevent memory-safety escapes. However, these techniques have limited usage in production deployments due to their performance and memory footprint impact as well as the need to rewrite some portions of code, incorporating Application Binary Interface (ABI) changes \cite{Google_MemorySafety}. In late 2025, the ChkTag feature was announced for the x86 architecture as a memory safety ISA extension with details of the architecture extension to become available later \cite{x86chktagIntel}.

In contrast, the ARM Architecture Memory Tagging Extension (MTE) introduced in ARMv8.5-A in 2018 \cite{ARMv85aIntroductionBlog}, \cite{ARM_MTE} is established as a promising architecture extension to detect/prevent memory safety escapes. The ISA extension allows associating tag values with pointers and physical memory. Based on comparison of the pointer tag value with the physical memory tag value for reads/writes, MTE enables deterministic detection and prevention of spatial sequential buffer overflow attacks and probabilistic detection and prevention of exploits resulting from temporal use-after-free pointer programming bugs \cite{liljestrand2022colorworlddeterministictagging}. Memory tag checking can typically be enabled without making any code changes to applications — at most needing to re-link the application with a tagging-enabled memory allocation library. This makes the feature attractive for use in production environments.

The AmpereOne$^\circledR$ processor, launched in 2024, is the first datacenter System-on-Chip (SoC) that provides support for ARM MTE \cite{AmpereOneM_Brief}. This paper provides details of its MTE implementation, which addresses the key memory capacity and performance overheads that have historically limited the deployment of memory-safety features. Through innovations in tag storage and core microarchitecture,  the AmpereOne$^\circledR$ processor delivers a hardware foundation designed to minimize the cost of tagging. Benchmarking results demonstrate that on this foundation, datacenter workloads can gain the benefits of MTE with only a single-digit percentage performance impact on a design that incurs no memory capacity overhead. This result establishes an efficient hardware baseline for what is fundamentally a hardware-software co-design effort. By mitigating the primary hardware costs, this work paves the way for the software ecosystem to further optimize allocators and runtimes, making performant, production-ready memory safety an achievable goal. 

This paper is structured as follows.  \Cref{sec:background} provides an architectural overview of the ARM Memory Tagging Extension. \Cref{sec:related} provides an overview of contemporary processor implementations of MTE targeted at handheld devices. Sections \ref{sec:challenges} and \ref{sec:taggingoptions} discuss aspects unique to a datacenter SoC MTE implementation and the tradeoffs among microarchitectural options.  \Cref{sec:implementation} provides details of the MTE implementation on the Ampere Computing AmpereOne$^\circledR$ processor. \Cref{sec:evaluation} provides details on the performance impact of software use of MTE for a range of datacenter workloads, along with an analysis of the cause of performance deltas. \Cref{sec:discussion} includes a discussion of potential future enhancements to the Ampere MTE implementation and implications of future memory technology directions. Finally, \Cref{sec:conclusion} provides concluding thoughts on MTE and the optimized implementation in the AmpereOne$^\circledR$ processor. 

\section{ARM Memory Tagging Extension Overview} % .5 pages
\label{sec:background}

% Reference doc: AmpereOne_Family_MTE_Supplement.docx
% Reference doc: AmpereOneXSKU-MemoryTagging-Customer-DeepDive

The ARM Memory Tagging Extension (MTE) is a memory-safety feature that provides mechanisms for the detection and prevention of the two main classes of memory-safety escapes. Specifically, with appropriate software use of tag values, it provides deterministic detection and prevention of sequential buffer overflow attacks and probabilistic detection and prevention of exploits resulting from  use-after-free and similar pointer programming bugs \cite{ARM_MTE}. MTE provides the above capabilities with the following architecture elements:
\begin{itemize}
    \item Physically addressable memory is partitioned into 16-byte \emph{granules}. Software (typically the \texttt{malloc} implementation) assigns a 4-bit \emph{allocation tag} to each granule using new tagging instructions. The architecture does not specify where allocation tags must be stored and leaves that choice to the SoC implementation.
    \item The pointer used to access the memory contains a 4-bit \emph{address tag} located in the upper unused bits of the virtual address. The address tag is programmed by software (e.g. \texttt{malloc}) when a pointer to allocated memory is returned. The ARM architecture provides a top-byte-ignore capability (TBI) \cite{ARMARM_TBI} to indicate that the highest order byte - in which the address tag is located - not be used by the processor when performing address translation for the virtual address.
    \item The address tag is checked against the allocation tag before every memory read/write access. The architecture allows for mismatches between the allocation and address tags to be handled either \emph{synchronously} or \emph{asynchronously}.
    \item The synchronous mode (SYNC) signals an exception on any mismatch and prevents the access from completing, similar to a page or permission fault.
    \item The asynchronous mode (ASYNC) allows the memory read/write to complete and sets a system register bit to indicate that a tag mismatch has occurred for later examination by software. 
\end{itemize}

\begin{figure}[!t] % for floating figures, use !t.  for inline, use H.
\centering
\includegraphics[width=\columnwidth]{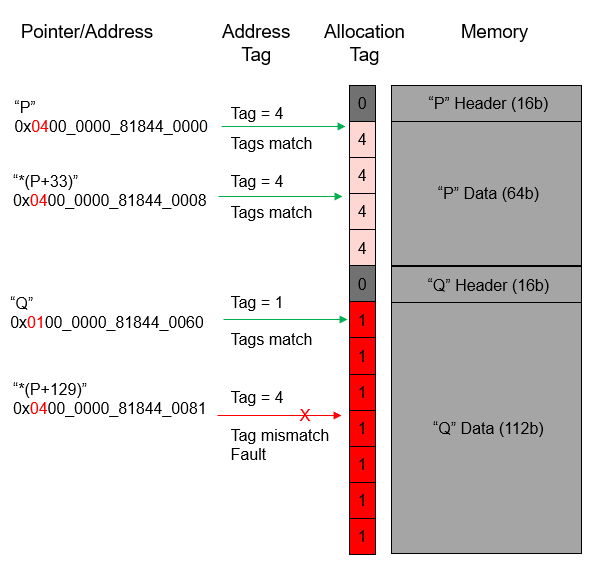}
\caption{Buffer Overflow detection and mitigation with Memory Tagging} 
\label{MemoryTaggingExample}
\end{figure}

With a careful choice of allocation tag values for buffers located sequentially in memory, MTE can detect, and in the synchronous mode, prevent buffer overflow accesses. Given that 4-bits are available for a tag value, memory allocation software reuses tag values resulting in the probabilistic detection and (in synchronous mode) prevention of escapes resulting from pointer programming/temporal use-after-free bugs. \autoref{MemoryTaggingExample} illustrates a scenario where use of memory tagging can effectively detect and mitigate a buffer overflow. In the example, there are two separate but sequentially located allocations of memory from a heap - the first returned with a pointer/address P and the second with a pointer/address Q. Memory allocated for pointer P is tagged with allocation tag value of 4 and the tag value 4 is programmed in bits 56:59 of the pointer virtual address as the address tag. Similarly, memory allocated for pointer Q is tagged with an allocation tag value of 1 and the tag value 1 is programmed in bits 56:59 of the pointer virtual address as the address tag. References from pointer P and pointer Q within their allocated memory ranges "pass" since the address tag and the allocation tags match. However, for an access using Pointer P that goes past the end of its allocated memory range into the memory range for allocation Q, the address tag value of 4 does not match the allocation tag value of 1. In SYNC mode, the PE raises a tag mismatch fault and the access does not complete. 

Microsoft published an analysis of the effectiveness of the deterministic and probabilistic protection provided by MTE for memory safety escapes reported in their CVEs \cite{MSR_tagging_analysis}. The analysis supports the underlying value proposition for MTE, providing:
\begin{itemize}
    \item Deterministic and durable protection for \textasciitilde{}13\% of memory-safety escapes reported (in the period covered by the report) associated with "Heap adjacent over-run/over-read" exploits.
    \item Probabilistic protection for "use-after free" bugs (\textasciitilde{}26\% of vulnerabilities reported in the period covered by the report) with attack probability reduced to 6\% if all tag bits are used, and
    \item Probabilistic protection for "Heap out-of-bounds read or write (non-adjacent)" bugs (\textasciitilde{}27\% of vulnerabilities reported in the period covered by the report) with attack probability reduced to 6\% if all tag bits are used.
\end{itemize}   

\section{ARM MTE in Mobile Processors} % 0.5 pages
\label{sec:related}

Since the introduction of the ARM MTE architecture extension in 2018, several processors for handheld devices with support for MTE have been brought to market. 

The Google Pixel 8 and Pixel 8 Pro \cite{GooglePixel8ProductBlog} phones, launched in October 2023, introduced support for ARM MTE with both synchronous and asynchronous tag-checking modes. MTE support is not enabled  by default on the production phone software, but is available as a capability to application developers. MTE in SYNC mode has successfully found use-after-free pointer bugs on these devices \cite{GoogleZeroPixel8MTE, ARMPixel8MTE}.

The Apple iPhone 17 and iPhone 17 Pro built on the Apple A19 and A19 Pro processors, respectively, and launched in September 2025, provide comprehensive support for the ARM MTE architecture extension \cite{AppleMemoryIntegrityEnforcementBlog}. Leveraging MTE support in the Apple A19 and A19 Pro processors, in conjunction with an extensive software investment in type-aware secure memory allocators and significant micro-architecture investments in the synchronous mode for MTE, the iPhone 17 and iPhone 17 Pro provide a comprehensive, always-on protection against attacks exploiting memory-safety escapes for the kernel and key user-mode processes. The use of the MTE SYNC mode is specifically called out as a requirement for effective protection from memory safety escapes. 

Both the Google Pixel 8 and Apple A19 processors have MTE implementations purpose built for handheld devices. Meanwhile, the datacenter platform, and software environment for which the MTE support in the AmpereOne$^\circledR$ processor is implemented, has unique differences from a handheld device's platform and software environment. The differences introduce a set of trade-offs, offering opportunities like the assurance of ECC-enabled memory configurations across all datacenter platforms, yet simultaneously posing challenges related to multi-tenancy and datacenter workload sensitivity to any memory bandwidth/latency degradation. Section~\ref{sec:challenges} covers details on these considerations when building MTE support in a datacenter targeted SoC - specifically the need for MTE SYNC mode for effective protection from memory safety escapes.

\section{MTE Considerations For datacenter SoCs} % 0.5 pages
\label{sec:challenges}

The datacenter platform and operating environment create unique challenges while providing opportunities to develop support for MTE in an SoC. This section outlines these along with a discussion of the implications for MTE support. 

% The needs of a datacenter-class CPU pose unique challenges for an MTE implementation. Performance is critical. Memory costs are high due to large memories and the need for ECC to support the high reliability expectations. Security is paramount, making it important to detect and report memory safety issues immediately. And the scale of cloud software creates a strong desire for extensive software debuggability. Servers must support diverse MTE usage models for the cloud users that share a server. All of this must be balanced in the implementation. The following paragraphs discuss each of these needs in detail.

\subsection{Memory Capacity and Reliability}
Financially, DRAM emerges as the most expensive ingredient in a datacenter platform. Meta's internal data reveals that DRAM's share of the total platform cost rose from 15\% in their first generation to a notable 35\% in their sixth generation platform~\cite{Meta_TMO}.

DRAM capacity is a key design point for datacenter platforms. Cloud Service Providers (CSPs) provide Platform-as-a-Service (PaaS) capabilities through Virtual Machines (VMs) that are configured for different workload classes with guarantees on the number of cores and amount of memory available to software in each VM profile \cite{Azure_VM, GoogleCloud_VM, OCI_VM}. Cloud service providers' datacenter platforms are designed to have the appropriate ratio of cores to total DRAM populated in the platform to support the optimal configuration of VM profiles from a Total Cost of Ownership (TCO) perspective. Any reduction of memory available to software has a direct impact on the density of provisioned VMs and the resulting cloud providers' TCO. 

For the MTE architecture, employing a 4-bit tag per 16 bytes of data, the memory footprint for storing allocation tag values is \textasciitilde{}3\% of the total software-visible memory ((4 bits / (16 bytes * 8 bits/byte + 4 bits)) = 3.03\%). The SoC implementation needs to provide memory for tag storage for the entirety of the memory accessible to software~\cite{ARMARM_TBI} given the potential for all of this memory to be tagged. Consequently, an implementation option that avoids reducing the capacity of software-available memory is highly desirable for MTE's production deployment. 

Datacenter environments are characterized by a high volume of platform deployments, each with large memory footprints. Given this scale, robust memory reliability features are paramount to detect and correct memory failures and ensure the necessary up-time for datacenter platforms. Thus, comprehensive ECC support across all DIMM configurations, including SECDED and SymbolECC\footnote{SymbolECC is a proprietary Ampere feature, based on the Reed-Solomon algorithm. Its correction properties are comparable to what is known in the industry as Chipkill (IBM), SDDC (Intel), AdvancedECC (HP), STAR ECC (Synopsys), and ExtendedECC (Sun Microsystems).} capabilities within the SoC, is considered a minimum requirement for deployed memory configurations \cite{MemoryRASDataCenter_IEEEAccess}.

\subsection{Multi-Tenancy}
Data shared by Microsoft indicates that more than 90\% of VM profiles supported in Azure datacenters are single vCPU, i.e., a single core is available to software running in the VM \cite{Azure_ResourceCentral_SOSP17}. In platforms with a high core count SoC such as the AmpereOne$^\circledR$ SoC with up to 192 cores, a very large number of VMs, each with different users/tenants, can be co-located and executing at the same time on the same platform. This multi-tenancy creates unique security challenges. The potential uniqueness of the software image in each VM creates a large surface area for memory-safety escapes. Furthermore, there is a risk that an exploit in one VM can expose other tenants on the same platform if the hypervisor and/or service operating system are exposed.  Hence, both detection and immediate prevention of attacks are necessary in production deployments. This implies that the SYNC mode for MTE must be used and the SoC should implement SYNC support in a performant fashion. Production cloud environments already provide strong isolation across VMs, and between VMs and the hypervisor, using a range of primitives including ISA based memory management and virtualization capabilities; so using memory tagging provides an additonal detection and mitigation capability for IOMMU containment for DMA traffic and Confidential Computing technologies such as SEV-SNP \cite{AMD_SEV_SNP}, TDX \cite{Intel_TDX} and ARM CCA \cite{ARM_CCA}. Also, there is recent growing interest in the use of memory tagging for isolation of un-trusted code from the rest of the system \cite{SFITAG_2023}. 

\begin{figure}[!t] % for floating figures, use !t.  for inline, use H.
\centering
\includegraphics[width=\columnwidth]{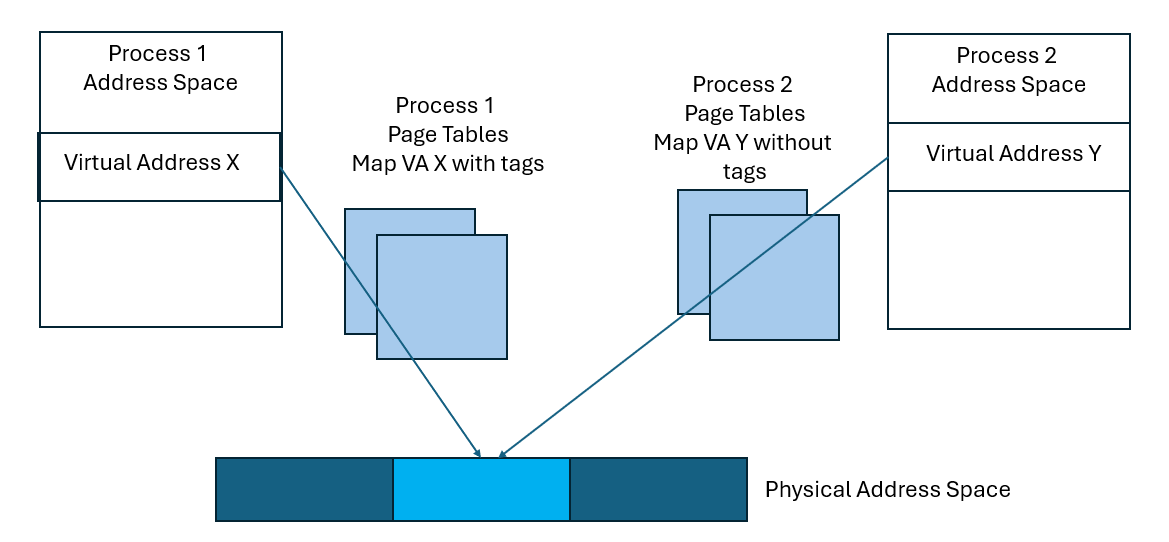}
\caption{Necessity for Tag Value preservation by the Processing Element.} 
\label{tagged_and_nontagged_mapping}
\end{figure}

Multi-tenancy introduces complexities concerning co-located tenants' use of the MTE feature. Unlike the homogeneous environment of a handheld device, where a phone provider has control over MTE usage across the software stack, a cloud environment would enable each tenant to independently configure MTE for their respective software images. From the SoC perspective, enabling MTE for even a single tenant requires the SoC to operate under the assumption that all allocation tags are potentially valid and must be preserved during memory writes. Although tagged memory pages are explicitly identified in page tables in the ARM MTE architecture, commercial operating systems allow for physical memory to be mapped by multiple page tables.  \autoref{tagged_and_nontagged_mapping} illustrates a scenario in which a physical address can be mapped into the address spaces of two different processes, one mapping with tagging enabled and the other without tagging. The first mapping at virtual address A in process 1 could be used to instantiate a valid set of tag values associated with the physical address space. A PE walking the page tables for virtual address Y in process 2 cannot assume that the target physical address does not have valid tag, even though the page table mapping indicates as such. Consequently, the SoC cannot solely rely on page table tagging attributes to determine tag preservation, and must conservatively maintain all tag values. This conservative approach can potentially introduce performance overheads impacting all co-located tenants. Given that various virtual machine profiles offered by cloud service providers include performance guarantees, it is critical for SoC implementations to minimize MTE-induced overhead, typically aiming for single-digit percentage performance impact.

\subsection{Goals for MTE support on the AmpereOne SoC}
In summary, the requirements of the datacenter operating environment motivated the following goals for the implementation of MTE on the AmpereOne$^\circledR$ SOC: 
\begin{itemize}
    \item Minimize or eliminate the memory capacity impact associated with MTE allocation tag storage without compromising memory reliability capabilities.
    \item Support the SYNC mode for tag checking as the primary usage mode, which is essential for production deployments. 
    \item Optimize the micro-architecture implementation to minimize performance impact in SYNC mode.
    \item Reduce performance degradation for software configurations that do not utilize tag checking, even when MTE is globally enabled on the SoC.
\end{itemize}

\section{Tagging Implementation Options and Tradeoffs} % 0.5 pages
\label{sec:taggingoptions}

Achieving these goals requires minimizing three categories of overhead:
\begin{itemize}
    \item \emph{Tag Storage Overhead}: Impact on available memory capacity due to the tag storage mechanism.
    \item \emph{Tag Fetch Overhead}: Latency introduced by retrieving allocation tags during memory read/write operations for tag checking.
    \item \emph{Tag Checking Overhead}: Computational cost associated with verifying the congruence between allocation and address tags.
\end{itemize}

\subsection{Tag Storage Considerations}

\begin{figure}[!t] % for floating figures, use !t.  for inline, use H.
\centering
\includegraphics[width=\columnwidth]{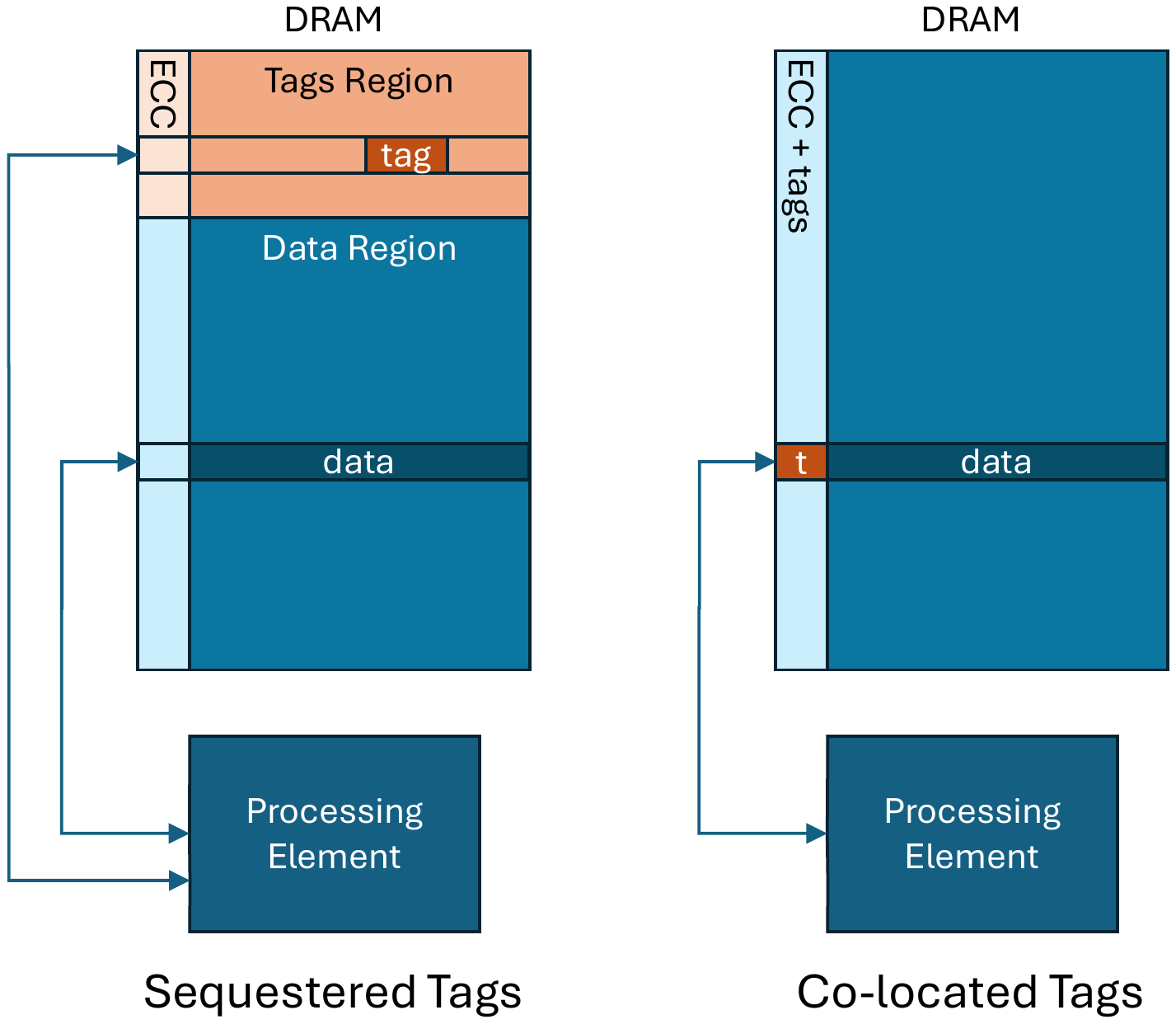}
\caption{Comparison of Tag Storage Methods} 
\label{tag_storage_methods}
\end{figure}

The organization of MTE allocation tags in memory is a critical design element because it determines the tag storage overhead, and therefore directly affects practicality of MTE deployments. Broadly, there are two ways to organize MTE allocation tags in memory.  These are depicted in \autoref{tag_storage_methods}.

\noindent \textbf{Statically allocate sequestered memory for tags:} 
This option involves reserving a dedicated portion of the total physical memory for tag storage at system boot. Given that each 16-byte memory granule requires four bits for its tag,  (4 bits / (16 bytes * 8 bits/byte + 4 bits)) = 3.03\% of the platform's physical memory must be exclusively reserved for tag data.
While methods for partitioning memory at boot are well-established for various specialized usages, their application for MTE tag storage presents significant challenges. As discussed in Section \ref{sec:challenges}, tag storage must be reserved for the entire physical memory present on the platform, due to the infeasibility of predicting which memory regions will be tagged by software at runtime. Consequently, sequestering 3\% of the total memory directly impacts the density of VMs that a CSP can provision, leading to an increase in TCO. As cloud servers are equipped with terabytes of installed DRAM, this 3\% reservation of memory translates into a substantial burden.

\noindent \textbf{Co-locate tag storage with data:} This option stores tags alongside the data in dedicated meta-data bits that are not otherwise usable as memory by software. This scheme eliminates the need to carve out software visible physical memory for tag storage up-front, and is favorable to datacenter deployments. 

In server platforms populated with memory technology that has ECC support, one option to store allocation tags is to use a subset of the ECC bits. Since these bits are used only by the memory controller unit (MCU) to provide memory reliability features, use of these bits does not impact total memory capacity available to software and eliminates the impact on VM density and TCO degradation to the cloud service provider. AmpereOne$^\circledR$ SoC servers use DDR5 registered DIMMs with ECC support providing access of 80 Byte Codewords (64B data + 16B of parity ECC), with a capability to correct a full symbol of 1 Byte. While this allows use of the ECC bits for tag storage, micro-architecture enhancements are needed in the MCU to provide the memory reliability capabilities needed in datacenter platforms using the remainder of the ECC bits available. This support is discussed later in Section~\ref{mte_in_mcu}. A similar technique for storing metadata in ECC bits is found in the Sun SPARC ADI implementation for Memory Tagging \cite{LinuxKernel_SPARC_ADI}, the RISC-V based Rocket SOC from lowRISC.org \cite{lowRISCorg_RocketSOC}, architectures for Confidential Computing \cite{cryptoeprint:2022/1472} and GPU memory safety \cite{sullivan2023implicit}.

\subsection{Tag Fetching Considerations}
\begin{figure}[!t] % for floating figures, use !t.  for inline, use H.
\centering
\includegraphics[width=\columnwidth]{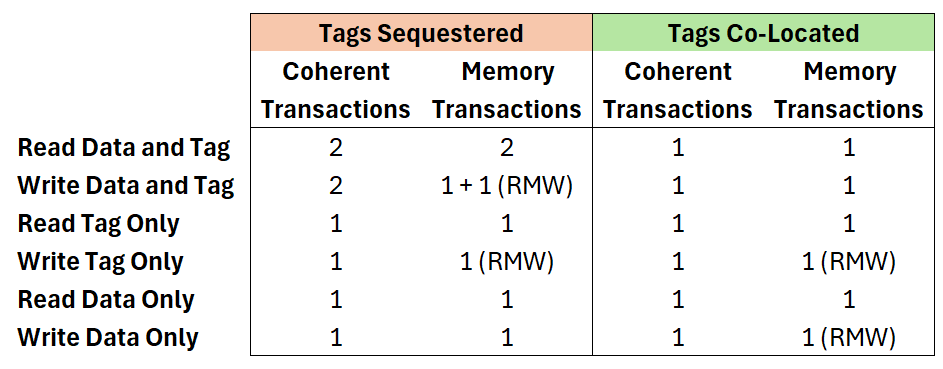}
    \caption{Transaction Counts per Operation}
    \label{tab:bwimpact}
\end{figure}

The organization of MTE tags in memory also affects the tag fetch overhead and hence, the run-time performance of applications that use MTE. If tags are not co-located with their data, the CPU read requests to tagged memory will require separate reads to fetch the corresponding allocation tags; these tag reads consume additional memory bandwidth and also potentially increase memory latency for read operations. CPU reads to untagged memory will not require additional memory reads but may be indirectly affected by the increased traffic due to tagged memory reads. Co-locating tag storage with data allows reading the tags along with the data without generating additional memory traffic and hence, doesn't negatively impact memory bandwidth and latency. \autoref{tab:bwimpact} shows the relative impact on bandwidth consumption of the MCU and Mesh interconnect on each core cache miss, based on the choice of tag storage method.  RMW indicates that the memory subsystem is required to perform a Read-Modify-Write to update the MTE tags. Previous work has demonstrated the performance impact of the increased traffic for memory tagging, in both CPU and GPU architectures \cite{lamster2024voodoo,sullivan2023implicit,cryptoeprint:2022/1472}.

\subsection{Tag Checking Considerations}

Synchronous tag checking can introduce latency to memory load and store operations, contingent on the specific implementation. This latency arises because tag validation must complete within the CPU pipeline before the memory operation is allowed to commit. This requires pipeline enhancements to perform the check without significantly degrading overall pipeline throughput.

When tags are sequestered from data, both loads and stores can experience additional latency, as tag retrieval requires separate memory transactions. In contrast, if tags are co-located with data, the tag check for load operations incurs virtually no overhead, as tags and data arrive simultaneously. Thus, co-location offers distinct advantages by impacting only store operations with added latency.

A notable disadvantage of co-location, however, is that untagged write-only transactions must be converted into read-modify-write (RMW) operations to preserve existing tags. This conversion can introduce latency and increase bandwidth requirements, particularly if the RMW operation entails loading the cache line into core caches. These overheads may be mitigated by the potentially higher bandwidth of the core to perform RMWs compared to the MCU.

\subsection{Summary}
An implementation where MTE allocation tags are co-located with data meets the requirements for MTE usage for production datacenter deployment:  no impact on usable memory capacity, lower hardware complexity, and lower performance overhead. MTE tag storage co-located with data can best be accomplished using ECC bits, but doing so requires enhancements to the memory reliability feature support built on ECC bits to continue providing the high level of memory reliability required for datacenter usages. See Section~\ref{mte_in_mcu} for more details.

\section{Ampere's MTE Implementation} % 2 pages
\label{sec:implementation}

The AmpereOne$^\circledR$ SoC, a high-performance datacenter server-class processor, offers configurations ranging from 96 to 192 custom Arm v8.6+ ISA-compliant Ampere cores. The AmpereOne$^\circledR$ SoC's MTE implementation incorporates enhancements across the Core Processing Element, including the L1 and L2 caches, the coherency mesh/interconnect, and the memory hierarchy, including the system-level cache and the MCU. This section details the MTE support integrated into each of these components, focusing on the design and micro-architectural choices that enable Ampere's efficient and performant implementation of synchronous tag checking.

\subsection{Overview}

\begin{figure}[!t] %  for floating figures, use !t.  for inline, use H.
\centering
\includegraphics[width=\columnwidth]{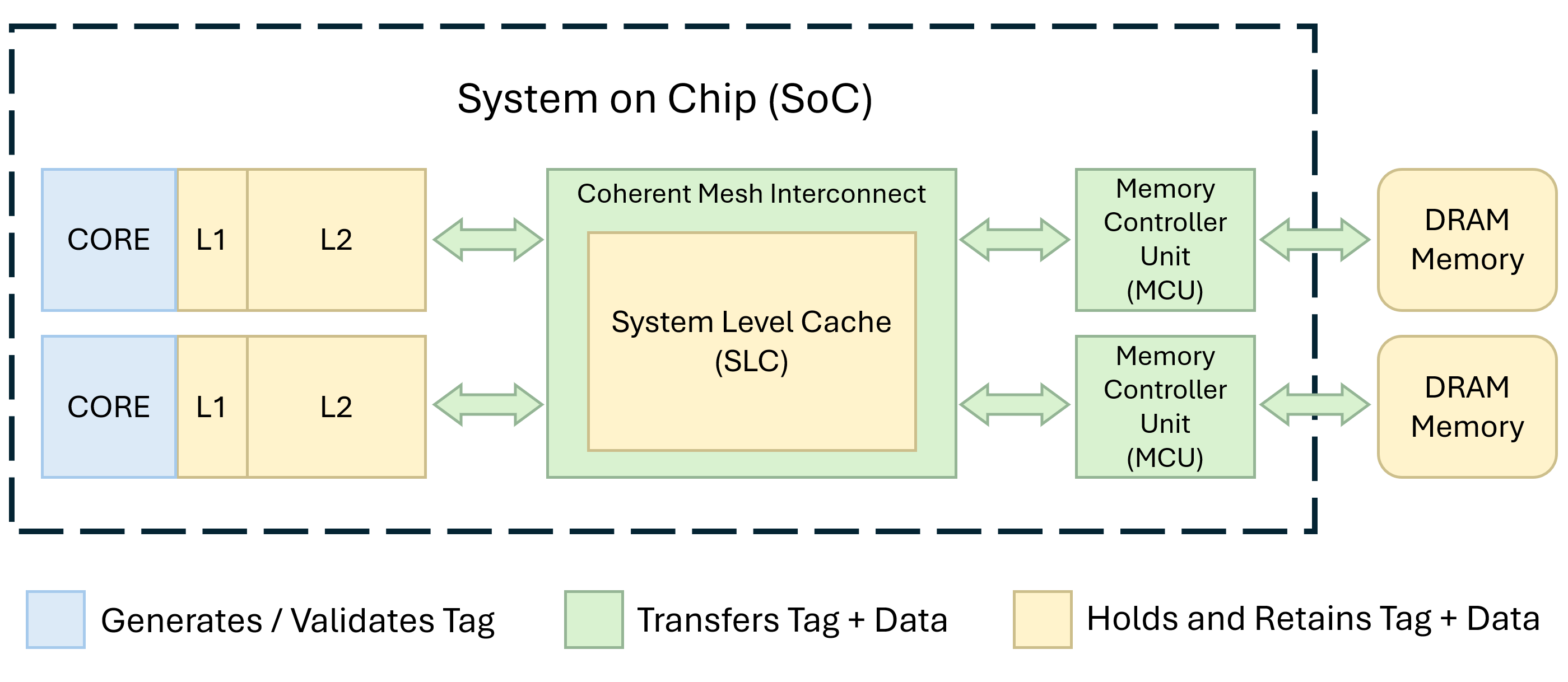}
\caption{Ampere MTE System Architectural View}
\label{fig:blockdiagram}
\end{figure}

In Ampere's implementation, tags and data flow together throughout the SoC, the Core Processing Element, and the memory subsystem. This is achieved by ensuring that every subsystem handles the tags and data as a co-located \emph{bundle}.  The bundle is first created at the memory subsystem during a memory read when the memory controller reads the tags that are stored in ECC bits inline with the data. The tags are  carried over the memory fabric using additional wires and held in caches in dedicated bits which flow through the Core Processing Element pipeline. \autoref{fig:blockdiagram} provides an overview of the SoC components and their roles in generating, retaining, and transporting tags.

\subsection{MTE in the MCU and DRAM}\label{mte_in_mcu}

Within the Memory Controller Unit (MCU), ECC bits are utilized to store allocation tags in DRAM. This necessitates the allocation of four ECC bits for every 16-byte DRAM granule. No tag checks are performed within either the MCU or the DRAM itself.

There is no latency impact within the MCU, as the tag bits are transferred along with the data bits, similar to Error-Correcting Code (ECC) bits. However, a notable impact arises concerning the utilization of ECC for error detection and correction. In certain DRAM and ECC configurations, insufficient metadata bits may be available to store MTE tags. To address this, some ECC bits must be re-purposed for tag storage. Ampere accommodates this by offering multiple ECC schemes tailored to various DRAM configurations. Specifically, a novel ECC scheme is employed that provides the requisite reliability protection for datacenter class SoCs while simultaneously making bits available for MTE allocation tags \cite{ecc2022}. The selection of the appropriate ECC scheme is determined by both the DIMM configuration and the desired MTE support. AmpereOne$^\circledR$ SoC systems support four schemes for ECC:
\begin{enumerate}
    \item SECDED baseline without MTE support (SECDED-64+8): The code word is 72 bits (64 bits data + 8 bits ECC) with 8 codewords per 64 byte cacheline. 
    \item SECDED with MTE support (SECDEC-128+4+9): The code word is 128+4+9 bits (128 bits data + 4 bits tag + 9 bits ECC) with 4 codewords per 64B cacheline. This scheme doubles the data granule from (A), allows storage of tag in spare unused ECC bits, retaining full memory reliability feature support without compromise.
    \item SymbolECC baseline without MTE support (SymbolECC-64+16): 64 bytes data divided into 8 codewords of 10 nibbles. Reed-Solomon allows one symbol correction per codeword. All the available metadata bits are used for ECC.
    \item SymbolECC with MTE support (SymbolECC-64-14+2): Similar to (C) but borrows 2 bits from ECC Parity, resulting in some Correctable errors becoming only Detectable.  Compared to (C), this provides 100\% of the fault detection, 100\% correction of bounded faults, and 99.98\% correction of unbounded faults.
\end{enumerate}    
For deploying MTE, options \textbf{B} and \textbf{D} are available to the cloud service provider, and both showcase reliability metrics within the bounds required for datacenter operation.

\subsection{MTE in the Coherent Mesh}

Within the mesh, tags are transported using reserved bits that are supported for implementation specific uses. This usage mirrors the use of metadata and ECC bits in the MCU and DRAM with tags being transported along with the data. On the AmpereOne$^\circledR$ SoC, this required customization of the mesh to allow for metadata transport along with data. All cache line storage in the mesh is expanded to include tag bits, and this storage is also covered by ECC. No tag checks are performed in the Mesh.

\subsection{MTE in the Core Processing Element (PE)}
Within the Core PE, data tags are transported alongside their corresponding data throughout the pipeline via widened datapaths and are stored co-resident with data in the widened caches. Tag validation is performed at the cache-lookup point within the PE pipeline. These checks proceed in parallel with address translation and access-permission checks (e.g., page-fault detection) and therefore do not introduce additional pipeline stages or stalls.

For loads, this design imposes negligible overhead relative to untagged execution because tag validation is integrated into the existing cache-lookup path. For stores, the primary effect is that the memory tag of the target cache line must be retrieved and validated before the store can commit, which constitutes the dominant MTE-related cost in the core. The implementation incorporates mitigations for this extra cost, including early line fetch for tagged stores.

For tagged stores, the cache-line fetch (including its tag) is initiated during address translation, just like for any normal load, which enables out-of-order overlap of the data/tag fetch with other work. As a result, tag check validation for stores can complete far earlier than the commit pipeline stage.

Store-to-load forwarding presents a specific challenge: a younger load may alias an older store whose target line’s memory tag has not yet been fetched and validated. To preserve correctness, the allocation tag (address tag) is recorded in the store buffer, and a load is permitted to forward only if its allocation tag matches that of the older store. At the time of forwarding, the outcome of the store’s tag check may still be unknown; however, if the allocation tags match, the subsequent tag validation will either succeed for both operations or fault the store (and thus the speculative load), maintaining correctness. Forwarding across a tag-store instruction (which writes the memory tag) is disallowed, as it could violate this invariant property. Thus, many stores experience no additional latency, while others incur a modest delay. 

\section{Evaluation of Ampere's MTE Implementation} % 2 pages
\label{sec:evaluation}

\subsection{Platform Configuration}\label{sec:platform}
The System Under Test (SUT) utilized an Ampere Computing reference platform, designated as Mt. Mitchell \cite{ampere_mtmitchell}. This rack server was equipped with an AmpereOne$^\circledR$ M SoC, featuring 192 cores operating at 3.2 GHz. Memory consisted of 512 GB of SK Hynix DDR5 modules, arranged in an 8x64 GB configuration and operating at 5200 MT/s. The configuration supported SymbolECC for RAS capabilities. For I/O, Samsung NVMe Solid State Drives (SSDs) provided storage, and Mellanox ConnectX 100GbE cards managed all network communications. 
To generate traffic, two 128-core Ampere Altra$^\circledR$ Max servers served as client systems. The clients are configured to keep the servers fully active under a P99 service level agreement. Typically, there is a one-to-one correspondence between client and SUT cores to saturate the system for performance measurement.

\subsection{Software Environment and Workloads}\label{sec:setup}

The Linux kernel has supported MTE since version 5.1.0 \cite{Linux_MTE}, with corresponding support added to GNU Binutils in version 2.45 \cite{GNU_Binutils} and to the GNU C Library (glibc) in version 2.33 \cite{Glib_MTE}. Fedora 36 and later distributions ship their kernels with CONFIG\_ARM64\_MTE enabled by default. For the following experiments, the system under test operated on Fedora 40 with GNU/Linux 6.10.6 and glibc 2.39.

To assess MTE's performance impact, motivated by similar evaluations \cite{utaustin_mte}, data was collected from a diverse range of real-world datacenter applications. These are industry-standard benchmarks, widely recognized for datacenter workload characterization and competitive performance evaluation. In all benchmark runs, all available cores on the SoC were fully utilized, to the extent each benchmark allowed. 

Memory tagging-specific overheads for tag fetching and checking are identical for workloads running bare-metal or within a VM; no additional VM-exits are introduced by tag setting or checking. We measured the MTE performance in a VM for subset of the workloads early in the AmpereOne program and found that the virtualization overhead (primarily from stage-2 translations) slightly dilutes the MTE performance impact compared to a bare-metal configuration. Hence, the paper focuses on the worse case of the two, bare-metal workloads, to fully expose the performance impact.

A summary of the benchmarks utilized is provided below.

    \para{memcached / memtier}\cite{memcached, memtier}:
        memcached is an open source in-memory NoSQL key/value distributed object caching database application. For performance evaluation, a client-server configuration was employed, using memtier\_benchmark-v1.3.0 load generators connected over a high-speed network to the server SUT. The SUT executed multiple independent instances of memcached-v1.6.21. 

        One important aspect of the SUT configuration was the allocation of CPU resources for IRQ handling. The Mellanox ConnectX NIC has 127 IRQs. If these network interrupts were handled by the same cores running memcached, it would lead to performance contention, particularly when running 192 memcached instances. This contention would result in significantly lower throughput and three times higher P99 latency. To mitigate this and adhere to latency SLAs, 20 SUT cores were explicitly dedicated to handling NIC IRQs, which left 172 cores to run memcached instances.

        The workload parameters were set to a 64-byte payload size and a 1:10 set-to-get ratio. To generate traffic, two Ampere Altra$^\circledR$ Max servers served as client systems, each configured to run 86 concurrent threads, totaling 172 client request threads. This setup aimed for a one-to-one correspondence between client threads and active SUT cores, for optimal system saturation indicative of a well-tuned cloud server.\footnote{The other cloud benchmarks' NIC and client setups were similarly tuned to achieve maximum server throughput in the baseline.}

    \para{Redis / memtier}\cite{redis, memtier}:
        Redis is an open~source in-memory NoSQL key/value data store that functions as either a database or an application cache. The experimental setup employed a client-server configuration: multiple client systems, running the memtier\_benchmark-v1.3.0 load generation tool, connected over a high-speed network to the System Under Test. The SUT executes multiple, independent, single-threaded instances of Redis-v7.2.0 servers. While Redis workloads typically leverage the jemalloc memory allocation library \cite{jemalloc_paper}, jemalloc currently does not support memory tagging. Therefore, to evaluate tagging impact, Redis was linked with glibc to use the standard malloc memory allocator.

    \para{vbench (H.264 Video Transcoding Benchmark)}\cite{vbench}:
        vbench is a benchmark designed for evaluating H.264 video transcoding scenarios in cloud-based video-as-a-service (VaaS) applications. It assesses transcoding performance relevant to typical cloud video workflows, and employs 15 H.264-compressed input files, varying in resolution (480p to 4K) and frame rates (25 to 60 fps). For this analysis, the focus was on two distinct VaaS profiles, Upload and Video On Demand (VOD):
        
    \emph{Upload Profile:}
        Measures transcoding speed and output quality for a first-uploaded temporary file. The primary objective is to make the video available for subsequent processing with minimal delay, without degrading the quality of the original input. This profile's reference uses a single-pass encoding with a constant quality target, allowing the encoder to use a high bitrate to maintain quality.
        
    \emph{VOD Profile:}
        Simulates scenarios where quality degradation impacts user experience. It strictly mandates that the transcoded quality must not be degraded compared to the reference. This profile's reference is based on an average case, employing a two-pass encoding strategy with a fixed bitrate target, processing in the background for VOD preparation.

    \para{nginx / wrk}\cite{nginx, wrkgit}:
        nginx is an open-source web server commonly employed as a reverse proxy, load balancer, or HTTP cache. To focus on server-side computational overhead, the SUT was configured with Brotli compression and the workload requests initiated server-side Lua scripts \cite{cloudflare}. Given this CPU-bound scenario, a one-to-one client-server configuration is employed where a single client running the wrk-v4.1.0 load generator was sufficient to saturate the SUT. The client connected via a high-speed  network to the SUT, which ran a single instance of nginx-v1.24 that spawned a worker process per core.

    \para{MySQL / sysbench}\cite{mysql, sysbench}:
        MySQL is an open~source, SQL-compliant, relational database management system (RDBMS). For performance evaluation Sysbench v1.1, a multi-threaded load generation and benchmarking tool, is employed. Sysbench was used to establish a simple database schema, populate database tables, and generate multi-threaded SQL query workloads. These queries were directed to a single instance of the MySQL-v8.0.43 database server, running concurrently with Sysbench on the SUT and communicating via TCP over a local network interface.

    \para{PostgreSQL / HammerDB}\cite{postgresql, hammerdb}:
        PostgreSQL is an open~source SQL-compliant, object-relational database management system (ORDBMS). For performance evaluation a client system running HammerDB-v4.3, a multi-threaded load generation and benchmarking tool to generate a TPC-C-like workload, was employed. This workload was transmitted over a high-speed network to the SUT, executing a single instance of the PostgreSQL v15.3 database server.

    \para{SPEC CPU$^\circledR$ 2017 integer base} \cite{cpu2017}:
        SPEC CPU was run in refrate mode with 192 copies to saturate the system and estimate SPECrate{$^\circledR$}2017\_int\_base. The binaries were built with open-source community gcc-15.1.1 with these base optimization flags: \texttt{-O3 -mcpu=ampere1a -flto}. 

\subsection{Performance Analysis}

Contemporary research \cite{utaustin_mte} has analyzed ARM MTE performance overheads using single-copy SPEC CPU, and a consolidated multi-tenant setup where both the server benchmark and client load generators are hosted on a single machine. In contrast, the measurements presented below are from 192-copy SPEC CPU and fully utilized server platforms with discrete client load generators. While  both approaches are valid, this one closely emulates real-world cloud deployments operating at peak utilization, and offers a unique yet complementary perspective on MTE performance. 

Three distinct system configurations are used for performance testing. Since MTE can be controlled at two levels, hardware (enabled/disabled in silicon) and software (tag checking enabled/disabled in user mode), this leads to the following three scenarios.
\begin{enumerate}
    \item MTE Disabled: The feature is disabled in silicon.
    \item MTE Enabled, No Tag Checking: MTE is enabled in silicon. The user-mode software does not use the tag checking enabled option for the glibc memory allocation library. 
    \item MTE Enabled, With Tag Checking: MTE is enabled in silicon. The user-mode software is configured to use the tag checking enabled option for the glibc memory allocation library. Tag checking is enabled by setting the environment variable for SYNC mode: \texttt{GLIBC\_TUNABLES=glibc.mem.tagging=3}
\end{enumerate}
This framework enables the performance analysis of two critical decisions: one for datacenter operators regarding hardware MTE enablement, and another for applications selecting to utilize MTE's memory safety features. In multi-tenant environments like cloud platforms, enabling MTE in hardware for one application or virtual machine could impose an inherent overhead on other co-located applications, even if they do not utilize MTE's tag checking. This ``always-on'' hardware overhead (\textbf{B} vs. \textbf{A}) is a critical consideration for cloud service providers. For an application intending to leverage MTE's safety features, the relevant overhead is the performance degradation when MTE with tag checking is fully active versus when MTE is available but tag checking is disabled (\textbf{C} vs. \textbf{B}).

\begin{figure}[!th] % 
\centering
\includegraphics[width=\columnwidth]{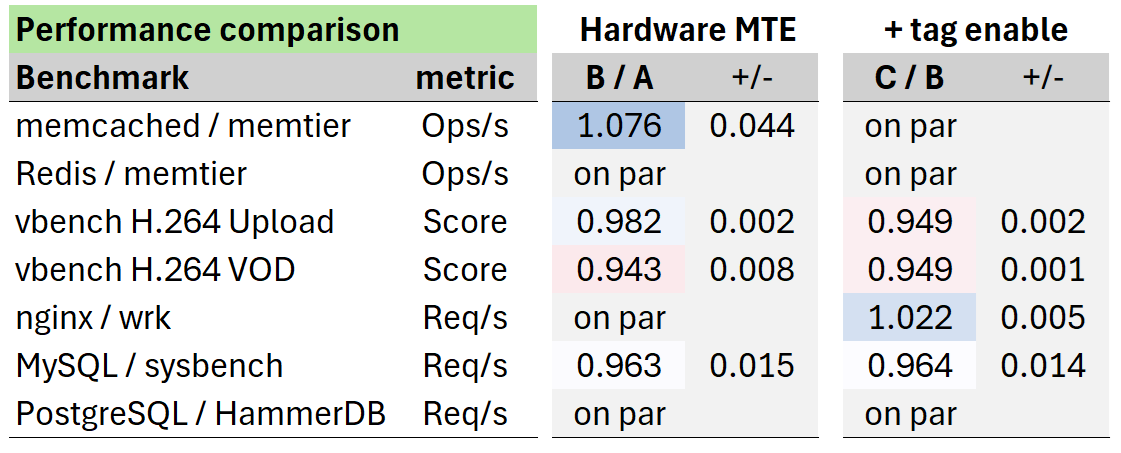}
    \caption{Performance impact on datacenter workloads. B/A shows the performance ratio when enabling MTE hardware, and C/B is the subsequent ratio after enabling tag checking. Each result is from five runs processed through \texttt{ministat}; ``on par'' indicates no difference proven with 99\% confidence.}
    \label{tab:perfdelta}
\end{figure}

\begin{figure}[!t] 
\centering
\includegraphics[width=\columnwidth]{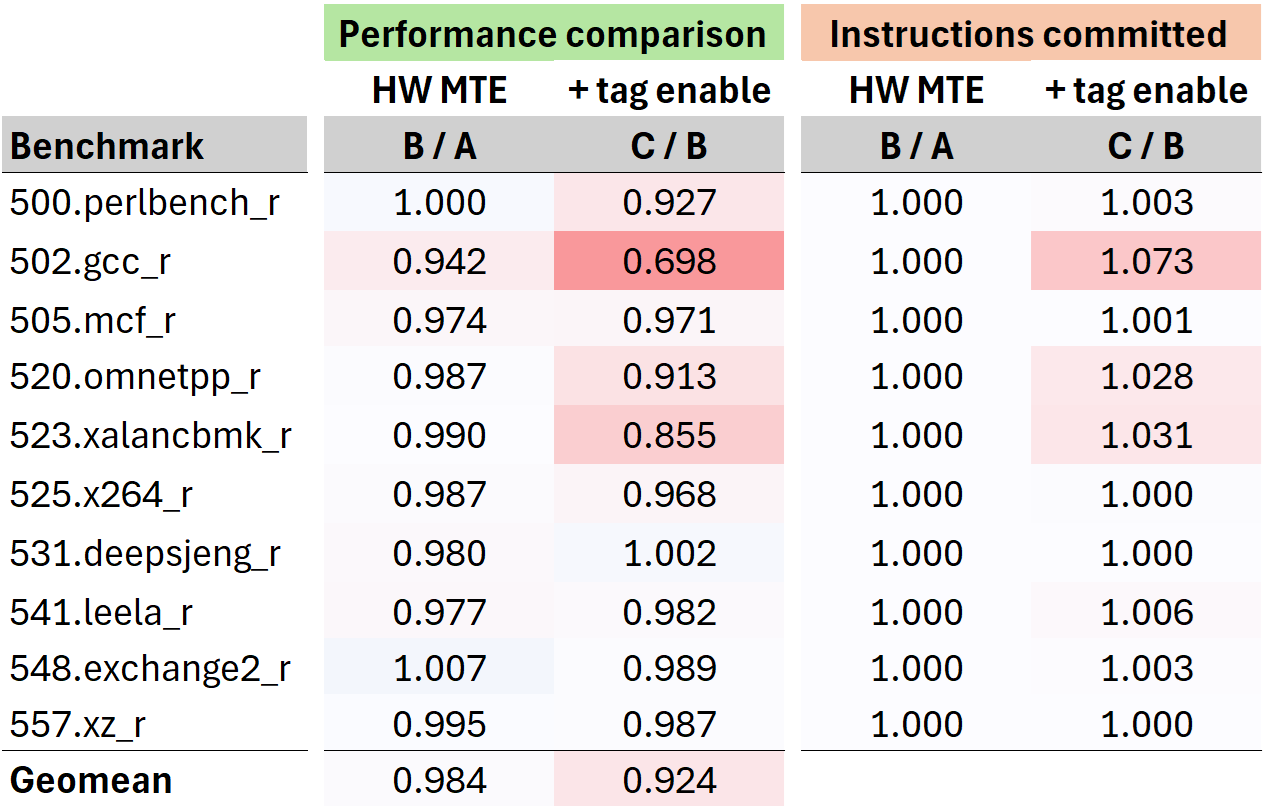}
    \caption{Performance impact on 192-copy Est. SPEC CPU$^\circledR$ 2017 
    benchmark scores. B/A shows the performance ratio when enabling MTE hardware, and C/B is the subsequent ratio after enabling tag checking. Analysis shows that committed instructions can increase when requesting tag checking.}
    \label{tab:cpu2017}
\end{figure}

To guide these two decisions, both performance ratios across the benchmarks are shown. \autoref{tab:perfdelta} presents the performance ratios and  measurement metrics for each datacenter cloud workload. \autoref{tab:cpu2017} reports the ratios for each SPEC CPU 2017 benchmark in 192-copy refrate.  To account for the inherent run-to-run variability observed in multi-threaded cloud workloads, each benchmark was executed five times. The resulting data were statistically analyzed using \texttt{ministat} \cite{ministat}, which compares two small populations to compute performance ratios and their associated error bars using Student's \emph{t}-test \cite{studentt}. All statistical comparisons were performed at a 99\% confidence level. Results designated as ``on par'' indicate that no statistically significant performance difference could be discerned between the compared populations. 

When looking at the \textbf{B} vs. \textbf{A} comparisons, enabling MTE in hardware can introduce a small overhead even when applications are not using tagging, due to SoC mechanisms that preserve tag state through the memory hierarchy. This platform-level comparison is relevant for multi-tenant deployments where a cloud provider enables MTE across the fleet. Across the workloads evaluated in Figures~\ref{tab:perfdelta} and \ref{tab:cpu2017}, that overhead was in the 1--6\% range, while memcached exhibited a 7\% improvement (with high variance). In this case, tag-loads effectively become prefetches, since tag-checking is turned off. Since tags and data are co-located, cache lines are brought in to the L1 data cache which happen to be beneficial to the instruction stream being executed. The source of the MTE performance benefit to memcached was confirmed with PMU data, which showed a significant reduction in time spent waiting for L2 misses, without a significant change in other operations which can prefetch lines into the cache, such as hardware prefetches or wrong-path speculation. Practically, these numbers suggest that tenants not asking for tag checking experience only modest impact when the platform enables MTE support for other users. 

Next, the comparison of \textbf{C} vs. \textbf{B}, which is the cost of turning on tag checking for applications. The overall performance impact of tagging is in the mid single-digit percentages with high confidence. Interestingly, in nginx shows a small but measurable speedup. This gain can be attributed to a prefetch-like effect: to perform tag checking, the cache line for a store is fetched earlier in the pipeline (prior to commit), allowing dependent memory operations to benefit from the prefetched line. When execution is limited by store latency and there is headroom in load-side bandwidth, this earlier fetch reduces the critical path and can improve overall throughput. The L2 prefetchers also get trained by tag-loads, which can lead to further benefits.

The primary sources of hardware slowdown stem from first-generation MTE integration in the AmpereOne$^\circledR$ microarchitecture and pipelines. First, as also observed by \cite{utaustin_mte}, store-to-load forwarding opportunities are reduced for tag-checked stores in the core. Second, the additional tag-read traffic increases pressure on L1D cache read ports, introducing structural hazards in the load/store pipeline. Refinements to address both of these issues have been incorporated into the next generation of Ampere's CPU cores.

Enabling MTE on the SPEC CPU 2017 suite results in a geometric mean performance degradation of 7.6\% on SPECrate, as detailed in \autoref{tab:cpu2017}. This figure is heavily influenced by significant regressions in 502.gcc and 523.xalanc, and to a lesser extent, 500.perlbench and 520.omnetpp. Our analysis of these four workloads indicates that the slowdown stems not from the hardware core issues cited above, but rather a combination of the following software factors:

\para{Increased Instructions Committed:}
The right-most columns in \autoref{tab:cpu2017} show that the number of retired instructions increases when memory tagging is enabled. This growth is attributable to two principal sources.

First, at user level, \texttt{malloc(3)} and related allocator paths in glibc expand to manage tag initialization and maintenance, adding instructions in the hot paths of allocation and deallocation to set allocation tags on newly returned memory and to clear tags on free. This includes the function \texttt{libc\_mtag\_tag\_region()} in glibc.

Second, at the kernel interface, glibc disables use of \texttt{brk(2)} for its arenas when tagging is requested and instead relies on \texttt{mmap(2)}. The \texttt{brk(2)} path, which primarily advances the program break to grow the heap (a simple move of a top-of-heap pointer), does not readily provide the page attributes or initialization required for MTE such as setting \texttt{PROT\_MTE} and preparing tag storage. In contrast, \texttt{mmap(2)} allows the allocator to request appropriately protected anonymous mappings, but it incurs higher instruction overhead: selecting a suitable virtual-address range, creating or merging VMAs (Virtual Memory Areas), updating page tables, honoring protection and flags, and triggering first-touch work and tag initialization. The kernel marks these pages as MTE-capable, and tag setup further increases instruction count and latency on first access.

\begin{figure}[!th]
\centering
\includegraphics[width=\columnwidth]{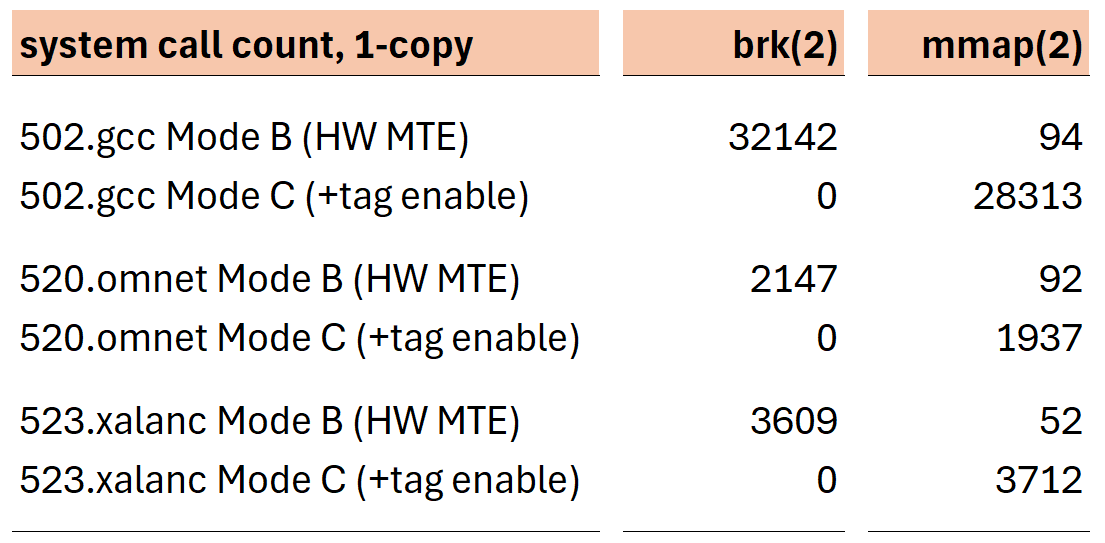}
    \caption{Breakdown of memory allocation system calls on the SPEC CPU benchmarks which show an increase in instructions committed in Mode C.} 
    \label{tab:syscalls}
\end{figure}

\autoref{tab:syscalls} illustrates this shift: with tagging enabled, calls that would have used \texttt{brk(2)} are replaced by \texttt{mmap(2)} based arena growth, increasing syscall and kernel work on the allocation path.

\para{Eager Tag Initialization:} 
Our investigation uncovered that 502.gcc and 523.xalanc allocate substantially large virtual address ranges while touching only a fraction of the space. Conventional allocators defer physical-page instantiation until first write; however, when an application requests MTE-tagged regions, these allocators perform eager tag initialization across the entire region before returning. This removes the benefits of lazy population and inflates first-touch costs, disproportionately affecting these workloads. The effect is amplified under homogeneous SPECrate runs, where identical phases execute in lock-step across cores, compounding the adverse behavior.

\begin{figure}[!th] 
\centering
\includegraphics[width=\columnwidth]{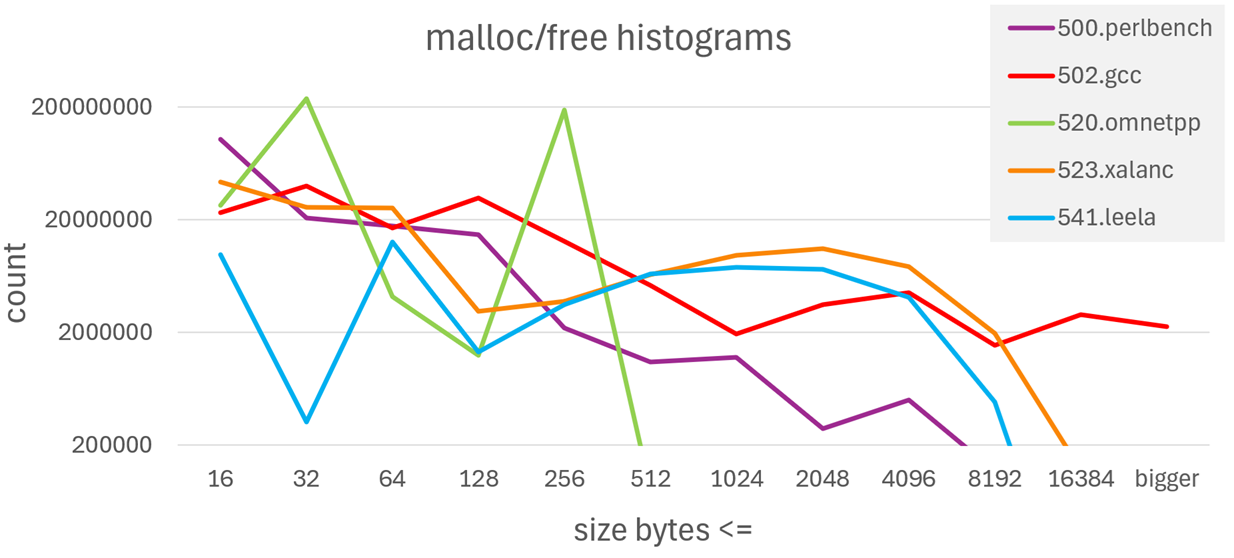}
    \caption{Distribution of memory allocation sizes and frequencies (log scale) for key benchmarks in SPEC CPU (1-copy). The workloads exhibiting the highest rates of small- to medium-sized allocations directly correspond to those most impacted by MTE overhead. This plot visualizes the allocation-intensive behavior that drives MTE performance degradation.}
    \label{tab:mallochistogram}
\end{figure}

\para{Transient Small-Object Allocations:}
The increased instruction execution detailed previously is not uniform; its magnitude depends heavily on an application's memory allocation patterns, specifically the frequency, size, and spatial locality of its allocations. The overhead is most acute in workloads characterized by frequent, small, and transient object allocations that result in a fragmented memory layout.
With MTE enabled, each call to malloc incurs a semi-fixed instruction cost to initialize the allocation tag. While this per-call overhead is easily amortized over large, long-lived allocations, it becomes a dominant factor for workloads that perform millions of small, short-lived allocations, directly increasing the user-space instruction count. To validate this hypothesis, we intercepted memory management calls using LD\_PRELOAD to profile their frequency and size. \autoref{tab:mallochistogram} plots the counts on a log scale for the five benchmarks exhibiting the largest increase in retired instructions. A strong correlation is evident: workloads with the most significant performance degradation are precisely those with the highest frequency of small-object allocations: 502.gcc, 520.omnetpp, 523.xalanc, followed by 500.perlbench. The other five benchmarks in CPU2017 have allocation counts too low to be visible on the chart's scale.

However, allocation frequency alone is insufficient to explain the full performance impact. For example, 500.perlbench also shows high allocation frequency but experiences a more modest slowdown. This suggests that the spatial pattern of allocations is a critical second factor. We hypothesize that workloads creating a fragmented virtual address space see amplified overheads at both the kernel and hardware levels. A fragmented layout increases the number of distinct VMAs the kernel must manage, making \texttt{mmap(2)} calls and page fault handling more expensive. This degrades spatial locality, leading to more TLB misses and costly page table walks.

This hypothesis aligns with the observed behavior. Workloads like 502.gcc, which construct complex, graph-like data structures (e.g., abstract syntax trees), naturally produce a more disjoint and fragmented allocation pattern. This fragmentation is exacerbated by 502.gcc's frequent use of \texttt{realloc(3)}. When these reallocations require moving data, their cost is amplified under MTE: the old memory region must be de-tagged and the new region must be tagged, effectively doubling the tag management overhead for a single high-level operation. In contrast, the more linear, stream-like processing in 500.perlbench with zero realloc's results in better memory locality and a less fragmented VMA layout, mitigating some of the kernel- and hardware-level penalties despite also having a high allocation rate of small objects. This finding is consistent with studies of other hardware-based memory safety features; prior research \cite{kim_heap_safety} similarly identified the high frequency of heap operations in 502.gcc and 520.omnetpp as the primary performance limiter due to the overhead of extra operations added to the allocator's critical path.

Therefore, the most significant MTE-related slowdowns occur when a high frequency of small transient allocations is combined with a fragmented access pattern within a large memory footprint. This scenario creates a ``perfect storm'' that stresses many sources of overhead simultaneously: the user-space allocator, kernel memory management, reallocation of memory, and the hardware's address translation mechanisms.

\begin{figure}[!ht] 
\centering
\includegraphics[width=\columnwidth]{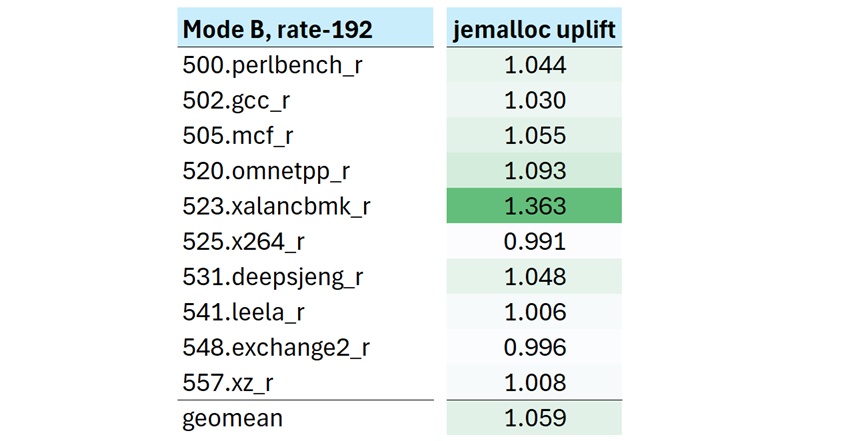}
    \caption{Performance improvement from linking the \texttt{jemalloc} allocator with SPECrate CPU$^\circledR$ 2017 benchmarks, running with 192 copies. The significant gains in allocation-intensive workloads provide evidence that MTE performance degradation can be caused by memory management overheads.}
    \label{tab:jemalloc_uplift}
\end{figure}

To further validate the hypothesis that frequent, fragmented allocations are a primary source of overhead, we evaluated the performance of these workloads with jemalloc \cite{jemalloc_paper}, a memory allocator designed to mitigate fragmentation. It accomplishes this through techniques such as size-classed arenas, which are particularly effective for workloads with high rates of small-object allocations. If our hypothesis is correct, the applications impacted by MTE should, in turn, derive benefit from a fragmentation-aware allocator like jemalloc.

Our results confirm this correlation. In \autoref{tab:jemalloc_uplift}, we show jemalloc's performance improvement on 192-copy CPU 2017, with a tagging-disabled baseline (since jemalloc does not support tagging). The workloads that exhibited the largest MTE-related regressions (502.gcc, 523.xalanc, and 520.omnetpp) do gain substantial performance when linked and run with jemalloc. This finding provides strong evidence that the observed MTE slowdowns are not an intrinsic cost of tag checking itself, but are tightly coupled to the underlying memory allocation patterns and the efficiency of the memory management subsystem. This study provides motivation for the community to research custom heap allocators with tagging support and smarter policies, which can additionally assist with other types of security issues \cite{alt_alloc, scudo}. Using type-based allocators or size-based allocators also aligns with Apple's direction of using a custom allocator for MTE, xzone malloc \cite{xzone_malloc}, where small blocks' tags are reassigned on free but larger blocks' tags are reassigned lazily upon the next allocation \cite{acm_queue_safety}. The GNU Tools Cauldron 2025 sessions also acknowledged the issue with MTE and GLIBC malloc of small objects \cite{glibc_cauldron}.

\section{Discussion and Future Work} % .25 pages
\label{sec:discussion}

This section outlines future opportunities for improving the performance and scalability of a data-center scale MTE implementation based on the AmpereOne$^\circledR$ SoC.

Co-location of tags with data by using a small number of ECC bits for tag storage has significant advantages for a datacenter SoC - no memory capacity impact and low performance overhead for a broad range of datacenter class workloads. However, as Sections \ref{sec:challenges} and \ref{sec:evaluation} point out, there are impacts to enabling MTE in the platform hardware, even if tag checking is not being used by software. For the AmpereOne$^\circledR$ SoC's ECC tag implementation, the small performance delta is due to the need to perform Read-Modify-Write operations when preserving tags for full-cacheline stores that would otherwise be Write-Only. Memory controller designs can consider optimizing this flow further to reduce total overhead. Also, as stated in Section \ref{sec:implementation}, the memory controller implementation in the AmpereOne$^\circledR$ SoC for option \emph{D} provides 100\% of the fault detection, 100\% correction of bounded faults, and 99.98\% correction of unbounded faults. As outlined in \cite{ecc2022} the "level of 9's" correction of unbounded faults can be increased with additional support in the memory controller implementation. Such improvements can be achieved by carrying error information forward from one codeword to one or more subsequent codewords in the same cacheline. By using this technique, which depends on the extreme unlikelihood of errors occurring in different DRAM devices in the same cacheline, the level of 9's can be improved significantly, to as high as 99.9999\%.

Some device classes with no knowledge of tags, such as PCIe devices, have the ability to directly access memory. While such accesses need to preserve allocation tags, doing so for an MTE implementation that co-locates tags with data has a very high degree of complexity for a performant implementation – with an extreme scenario involving read-modify-write transactions for each cache line with the associated significant impact on memory bandwidth. To deliver an optimal system-level solution, the AmpereOne$^\circledR$ SoC MTE implementation stipulates that software either not share SW memory intended to be utilized for tagging with devices, or preserve tags around these memory access operations. Looking forward, the ubiquitous presence of System Memory Management Units (SMMUs) in all device-to-memory paths within datacenter SoCs presents an opportunity for architectural advancement. Future SMMU enhancements could incorporate tag awareness, allowing the SMMU to identify and manage device accesses to non-tagged memory regions more optimally. Such an architectural improvement would allow microarchitectural optimizations for these device accesses, thereby eliminating the current software dependence.

The AmpereOne$^\circledR$ processor uses bits reserved in the mesh interconnect for metadata to transport tag values alongside the data. The ARM CHI-E extension added explicit support for tag management and transport in the mesh protocol \cite{AMBA_CHI_Spec}. Future Ampere devices that support the CHI-E mesh protocol will leverage the tag management support for transporting tag values with data using these protocol extensions.

Industry standards such as the Compute Express Link (CXL-3.1) provide profiles for CXL devices with memory, referred to as CXL.mem or CXL Type 3 configurations \cite{CXLSpec}. Enhancements to the specification for carrying metadata have been defined in the CXL 3.2 specification \cite{CXLSpec3_2}. Future MTE implementations leverage these extensions to support memory tagging for the CXL.Mem/Type 3 devices. The support requires commercial availability of memory technology devices that support the CXL 3.2 metadata extension along with sufficient storage for tags in the metadata bits.

Similar to the CXL 3.2 extension, it is possible that future DRAM technologies add support for metadata transport. If such standards emerge, future MTE implementations may benefit from leveraging the metadata support for tag transport. 

\section{Conclusion} 

\label{sec:conclusion}

Memory safety remains a major challenge for the large body of software written in pointer-based languages such as C/C++. While compiler and ISA extensions have been proposed to mitigate these risks, their adoption has been limited by memory capacity and performance overheads. The ARM Memory Tagging Extension (MTE) offers a practical path forward, providing robust, hardware-enforced detection of spatial and temporal memory errors. A key attraction of MTE is its potential for deployment by simply linking applications with a tagging-aware memory allocator. However, as demonstrated in this work, the largest performance overheads stem not from the core MTE mechanism itself, but from the interaction between application allocation patterns and the memory management subsystem.

The AmpereOne$^\circledR$ processor addresses the hardware side of this co-design challenge. It integrates system-wide support for ARM MTE while eliminating the memory capacity tax for tag storage and delivering single-digit percent overheads in synchronous mode for a range of datacenter workloads. This is enabled by a co-located data-and-tag design throughout the widened datapaths, caches that store tags alongside data, and core pipeline mechanisms for parallel tag checking -- which together provide a highly efficient hardware foundation for memory tagging.

Looking ahead, the path to ubiquitous, low-overhead memory tagging hinges on continued hardware-software co-design. While future hardware will continue to refine the microarchitecture and memory architecture, a significant and immediate opportunity now exists within the software ecosystem. The development of advanced, MTE-aware memory allocators that can mitigate fragmentation and more efficiently manage the lifecycle of tagged transient objects is the essential next step. By building on the performant hardware platform presented here, the software community can unlock the full potential of MTE, making broadly deployed, low-overhead memory safety a reality for production data-center environments.

\section*{Acknowledgement}

%\label{sec:acks}
This work would not have been possible without the contributions of many of our colleagues, including the designers and verification engineers that implemented and validated MTE. We would like to specifically thank our performance analysis team: Olivier Singla, Jiangning Liu, Feng Xue, James Burke, Sean Daniels, and Steve Winiecki for conducting extensive performance measurements, and Syed Fakhri for his support of the performance monitoring infrastructure.  We are also grateful to Richard Shannon, Mark Charney, Shahid Khan, and Shravan Narayan for their valuable guidance and reviews.

%\clearpage
\printbibliography

\end{document}